\documentclass[12pt]{article}
\usepackage{amsfonts}
\newcommand{\be}{\begin{equation}}
\newcommand{\ee}{\end{equation}}
\newcommand{\bea}{\begin{eqnarray}}
\newcommand{\eea}{\end{eqnarray}}
\newcommand{\non}{\nonumber}
\def\e{\epsilon}

\begin{document}

\begin{titlepage}
\begin{flushright}
hep-th/0505188
\end{flushright}
\vskip 1.5in
\begin{center}
{\bf\Large{$\bf R^2$ Corrections for 5D Black Holes and Rings}}
\vskip 0.5in {Monica Guica\footnote{Permanent
address:\textit{Jefferson Physical Laboratory, Harvard University,
Cambridge MA 02138, USA} }, Lisa Huang$^{1}$, Wei Li$^{1}$ and
Andrew Strominger$^{1}$} \vskip 0.3in {\small{ \textit{ Center of
Mathematical Sciences, Zhejiang University, Hangzhou 310027
China}}}

\end{center}
\vskip 0.5in

\baselineskip 16pt
\date{}

\begin{abstract}

We study higher-order corrections to two BPS solutions of 5D
supergravity, namely the supersymmetric black ring and the spinning
black hole. Due in part to our current relatively limited
understanding of F-type terms in 5D supergravity, the nature of
these corrections is less clear than that of their 4D cousins.
Effects of certain $R^2$ terms found in Calabi-Yau compactification
of M-theory are specifically considered. For the case of the black
ring, for which the microscopic origin of the entropy is generally
known, the corresponding higher order macroscopic correction to the
entropy is found to match a microscopic correction, while for the
spinning black hole the corrections are partially matched to those
of a 4D $D0-D2-D6$ black hole.

\end{abstract}
\end{titlepage}
\vfill\eject

\tableofcontents

\section{Introduction}

Recently a surprisingly powerful and precise relationship has
emerged between higher dimension F-terms in the 4D effective
action for ${\cal N}=2$ string theory (as captured by the
topological string \cite{bcov}) and the (indexed) BPS black hole
degeneracies \cite{dutch,osv}. Even more recently \cite{gsybh} a
precise relationship has been conjectured between the 4D and 5D
BPS black hole degeneracies. This suggests that there should be a
direct relationship between higher dimension terms in the 5D
effective action and 5D degeneracies which does not employ four
dimensions as an intermediate step. Five dimensions is in many
ways simpler than four so such a relation would be of great
interest. It is the purpose of this paper to investigate this
issue.

The 4D story benefitted from a well understood superspace
formulation \cite{mohaupt, dlvp}. The relevant
supersymmetry-protected terms are integrals of chiral superfields
over half of superspace and can be classified. In 5D the situation
is quite different (see e.g. \cite{berg}). There is no superfield
formulation and we do not have a general understanding of the
possible supersymmetry-protected terms. In general, the uplift to
5D of most of the 4D F-terms vanishes. However, the area law
cannot be the exact answer for the black hole entropy (for one
thing it doesn't give integer numbers of microstates!) so there
must be some kind of perturbative supergravity corrections.

As a first step towards a more general understanding, in this
paper we will study the leading order entropy correction arising
from $R^2$ terms, which are proportional to the 4D Euler density.
Such terms give the one loop corrections in 4D, and - unlike the
higher order terms - do not vanish upon uplift to 5D. They are
also of special interest as descendants of the interesting  11D
$R^4$ terms\cite{gutperl, chirita}. These terms correct the
entropy of the both the 5D black ring \cite{eemr} and the 5D BMPV
spinning black hole \cite{bmpv}. We find that the macroscopic
black ring correction matches, including the numerical
coefficient, a correction expected from the microscopic analysis
of \cite{cgms}. For the BMPV black hole, we find the correction
matches, to leading order, one expected from the 4D-5D relation
conjectured in \cite{gsybh}.

The next section derives the $R^2$ corrections to the 5D entropy
as horizon integrals of curvature components  using Wald's
formula. Section 3 evaluates this formula for the black ring,
while section 4 evaluates it for BMPV. Section 5 contains a brief
summary.

\section{Wald's formula in 5D}
In this section we will use Wald's formula to derive an expression
for $R^2$ corrections to the 5D entropy.

The Einstein-frame low energy effective action for the
compactification of M-theory on a Calabi-Yau threefold $CY_3$ down
to five dimensions contains the terms \cite{ferrara}

 \be I_0+\Delta I= - \frac{1}{32 \pi^2 } \int d^5 x\sqrt{|g_5|} R^{(5)} - \frac{1}{2^9\cdot 3 \pi^2 }\int
d^5x\sqrt{|g_5|}c_{2A}Y^A(R_{\alpha\beta\mu\nu}R^{\alpha\beta\mu\nu}
- 4 R_{\alpha\beta}R^{\alpha\beta} + R^2) \label{corr} \ee in
units in which $G_5=2\pi$ (for compactification on a circle of
unit radius, this choice leads to $G_4=1$ and hence facilitates
4D/5D comparisons). Here $Y^A$, $A=1,...n_V$ are scalar components
of vector multiplets. They are proportional to the K\"ahler moduli
of $CY_3$, normalized so that

\be D_{ABC} Y^A Y^B Y^C =1. \ee $c_{2A}$ are the components of the
second Chern class of $CY_3$ and $D_{ABC}$ the corresponding
intersection numbers. The $R^2$ term in  $\Delta I $ arises from
dimensional reduction of the much studied $R^4$ term
\cite{gutperl,chirita} in eleven dimensions.  It is also the
uplift from four dimensions of an $F$ term whose coefficient is
computed by the $N=2$ topological string on $CY_3$ at one loop
order \cite{bcov}.

When we add $R^2$ corrections to the action the entropy is no
longer given by the area law; instead, we need to use the more
general formula found by Wald \cite{waldref} \be  S_{BH} = 2\pi
\int_{Hor} \frac{\partial \mathcal{L}}{\partial
R_{\mu\nu\rho\sigma}} \e_{\mu\nu} \e_{\rho\sigma} \ee where,
$\epsilon_{\alpha\beta}$ is the binormal to the horizon, defined
as the exterior product of two null vectors normal to the horizon
and normalized so that $\e_{\alpha \beta} \e^{\alpha\beta} = -2$.
We can then identify two types of first-order corrections implied
by this formula:
\begin{itemize}
\item modifications to the area law due to the additional terms in
the action  - these terms are evaluated using the zeroth order
solutions for the metric and the other fields.

\item modification of the area due to the change of the metric on the horizon, which
follows from the fact that adding extra terms to the action may
change the equations of motion.
\end{itemize}

In 4D, the second type of modification is absent at leading order
for this particular $R^2$ form of $\Delta I $ obtained by
reduction of (\ref{corr}) \cite{msw}.   This and the 4D-5D
agreement we find to leading order suggest that this may be the
case in 5D as well. In order to understand all $R^2$  corrections
to the entropy this should be ascertained  by direct calculation.
In the following we consider only the first type of modification.

The corresponding correction to the entropy is then (see also
\cite{myers})

\be \Delta S = - \frac{4 \pi c_2 \cdot Y}{2^9 \cdot 3\pi^2}
\int_{Hor} d^3 x \sqrt{h} \left( \; R_{\mu\nu\rho\sigma}
\e^{\mu\nu} \e^{\rho\sigma} - 4 \;R_{\mu\rho} g_{\nu\sigma}
\e^{\mu\nu} \e^{\rho\sigma} +  \; R \e_{\mu\nu} \e^{\mu\nu}
\right) \ee where $h$ is the induced metric on the horizon and the
moduli are fixed at their attractor values. In the following we
will evaluate this correction for the spinning black hole and
black ring solutions.

\section{The black ring}

The black ring solution was discovered in \cite{eemr} and its
entropy understood from a microscopic perspective in \cite{cgms}. It
represents a supersymmetric solution to 5D supergravity coupled to a
number of abelian vector (and hyper)multiplets that describes a
charged, rotating black ring. It is characterized by electric
charges $q_A$, magnetic dipole charges $p^A$, and the angular
momentum around the ring, $J_{\psi}$. The macroscopic entropy
formula for the black ring can be written in the suggestive form

\be S_{BR} = 2 \pi \sqrt{\frac{c_L \hat{q}_0}{6}} \label{sbr} \ee
where, in terms of the macroscopic charges,

\be c_L = 6D = 6D_{ABC} p^A p^B p^C \label{ccharge} \ee $D_{ABC}$
being (one sixth) the intersection numbers of the Calabi-Yau, and

\be \hat{q}_0 = - J_{\psi} + \frac{1}{12} D^{AB} q_A q_B +
\frac{c_L}{24} \label{hatq} \ee where $D^{AB}$ is the inverse of
$D_{AB} \equiv D_{ABC} p^C $. The microscopic origin of the entropy
is from the quantum degeneracy of a 2D CFT with central charge $c_L$
and left-moving momentum $\hat q_0$ available for distribution among
the oscillators. The last term in (\ref{hatq}) is ascribed to the
left moving zero point energy.

\subsection{Macroscopic entropy correction}
Now we evaluate the correction to the black ring entropy induced by
$\Delta I $.  Due to the 5D attractor mechanism \cite{kallars} the
moduli take the horizon values

\be Y^A = \frac{p^A}{D^{\frac{1}{3}}} \ee Next, all we need to do is
to find the binormal to the horizon for the black ring metric,
evaluate the relevant curvature terms at the horizon, and integrate.
We obtain\footnote{In our computation we have employed the following
relationships between various quantities used in this paper, in
\cite{eemr} and in \cite{bmpv}:

$$ Q^{eemr} = (16 \pi G)^{\frac{2}{3}} \mu^{bmpv} =
\left(\frac{4G}{\pi} \right)^{\frac{2}{3}} q \label{conv}$$

$$ q^{eemr} = \left(\frac{4G}{\pi} \right)^{\frac{1}{3}} p  $$

$$ J = J_{eemr} = 16 \pi J_{bmpv} = 4 \pi^2 \mu \omega$$The
value of Newton's constant used in \cite{bmpv} is $G_5 =
(16\pi)^{-1}$, so we needed to rescale their metric by $(16\pi
G)^{\frac{2}{3}}$ in order to get ours. Also recall we are setting
$G=2\pi$ in the text. }

\be \Delta S_{BR} =  \frac{\pi}{6} \;  c_2 \cdot p \;
\sqrt{\frac{\hat{q}_0}{D}} \label{brent} \ee

\subsection{Microscopic entropy correction}
 The microscopic entropy comes from M5 branes wrapping 4 cycles associated to $p^A$ in
 $CY_3$.  As shown in \cite{msw},  these are described by a CFT with left-moving central
charge

\be c_L = 6 D + c_2 \cdot p \label{rrf} \ee In \cite{cgms} the
leading entropy at large charges was microscopically computed using
the leading approximation (\ref{ccharge}) to $c_L$ at large charges.
Subleading modifications should arise from using the exact formula
(\ref{rrf}) in (\ref{sbr}). This leads to

 \be \Delta S_{BR} = \frac{\pi}{6}\; c_2 \cdot p \; \sqrt{\frac{\hat{q}_0}{D}} +
\frac{\pi}{24}\; c_2 \cdot p \; \sqrt{\frac{D}{\hat{q}_0}} + \ldots
\ee The first term comes from correcting $c_L$ in (\ref{sbr}), while
the second comes from correcting the zero point shift in
(\ref{hatq}). We see that the macroscopic $R^2$ correction matches
precisely the first term. We do not understand the matching of the
second term, but note that it is subleading in the regime $\hat q_0
\gg D$ where Cardy's formula is valid.

\section{The BMPV black hole}
Let us now turn now to BMPV - the charged rotating black hole in 5D
characterized by electric charges $q_A$ and angular momenta $J$ in
$SU(2)_L$. Its leading macroscopic entropy is given by

\be S_{BMPV} = 2 \pi \sqrt{Q^3 - J^2} \ee where \be Q^{\frac{3}{2}}
= D_{ABC} y^A y^B y^C \ee where the $y^A$'s are determined
from

\be q_A = 3 D_{ABC} y^B y^C \ee We find the correction to this
entropy following from the application of Wald's formula to
(\ref{corr}) to be

\be \Delta S_{BMPV} = -\frac{\pi}{24} \sqrt{Q^3-J^2}  \; c_2 \cdot Y
\left( -\frac{3}{Q} - \frac{J^2}{Q^4} \right) = \frac{\pi}{6}
\mathcal{A} \; c_2 \cdot Y \left(\frac{1}{Q} -
\frac{\mathcal{A}^2}{4 Q^4}\right) \label{dels2}\ee where we defined
$\mathcal{A} = \sqrt{Q^3-J^2}$ and the moduli fields take the
horizon values\footnote{From now on we will take `$Y^A$' to mean the
horizon value of the modulus field $Y^A$.}

\be Y^A = \frac{y^A}{Q^{\frac{1}{2}}} \ee In general, the
microscopic origin of the entropy for the 5D spinning black holes
in M-theory on $CY_3$ (unlike for black rings) is not
known,\footnote{It is of course known for ${\cal N}=4$
compactifications \cite{sventropy,bmpv}, so it would be
interesting to interpret the macroscopic correction for that
case.} so we will not try herein to understand the microscopic
origin of $\Delta S_{BMPV}$. We will however compare it to
corresponding corrections in 4D and the topological string
partition function. As argued in \cite{gsybh}, the exact 5D BMPV
entropy is equal to the entropy of the $D6-D2-D0$ system in 4D,
with the same 2-brane charges $q_A$, $D6$-brane charge $p^0=1$,
and $D0$-brane charge $q_0 = 2 J$. In the same paper, the
following relationship for the partition functions of 5D black
holes, 4D black holes and consequently of the topological string -
see \cite{osv} -was conjectured

\be Z_{5D} (\phi^A, \mu) = Z_{4D} (\phi^A , \phi^0 = \frac{\mu}{2} +
i \pi ) = \left| Z_{top}\left( g_{top} = \frac{8 \pi^2}{\mu}, t^A =
\frac{2 \phi^A}{\mu} \right) \right|^2 \ee where $\phi^A$ are the
electric potentials conjugate to $q_A$, while $\phi^0$ is conjugate
to $q_0$ in 4D, and $Re \mu = ( \mu + 2\pi i)$ to $J$ in 5D. The
absolute value in the last expression is defined by keeping $\phi^0$
real. With this in mind, we can start from $F_{top}$ - the
topological string amplitude- and compute the entropy of the BMPV
(including first order corrections) as follows. Up to one-loop order
$F_{top}$ is

\bea F_{top} & = &  \frac{i (2\pi)^3 }{g_{top}^2} D_{ABC} t^A t^B
t^C - \frac{ i \pi}{12} c_{2A} t^A \\ \non & = &  \frac{i}{\pi}
\frac{D_{ABC}\phi^A \phi^B \phi^C}{\mu} - \frac{i \pi}{6}
\frac{c_{2A} \phi^A}{\mu}\eea The entropy of the black hole is
given by the Legendre transform of

\be \mathcal{F} (\phi^A, Re \mu) = \ln Z_{BH} = F_{top} +
\bar{F}_{top} \ee To first order we have

\be \mathcal{F} = -\frac{1}{\pi^2} \frac{D_{ABC}\phi^A \phi^B
\phi^C-\frac{\;\,\pi^2}{6}c_{2A}\phi^A}{(\frac{Re \mu}{2\pi})^2
+1} \ee which gives

\bea q_A & = & \frac{1}{\pi^2} \frac{3 D_{ABC} \phi^B \phi^C -
\frac{\;\,\pi^2}{6} c_{2A}}{(\frac{Re \mu}{2\pi})^2 +1} \\ \non
 J & = & - \frac{Re \mu}{2 \pi^4} \;  \frac{D_{ABC}\phi^A \phi^B
\phi^C - \frac{\;\,\pi^2}{6} c_{2A}\phi^A}{\left((\frac{Re
\mu}{2\pi})^2+1\right)^2} \label{legend}\eea and therefore

\be S = 2 \pi \sqrt{Q^3 -J^2} (1+ \frac{1}{12}  \frac{c_{2A}
Y^A}{Q} + \ldots ) \label{bhentropy} \ee where the $\ldots$ stand
for higher order corrections in $|g_{top}|^2 = 16 \pi^2
\mathcal{A}^2/Q^3$.

We see that to the 5D $R^2$ corrections (\ref{dels2}) to the
entropy do not exactly match the 4D corrections (\ref{bhentropy}).
This is possible of course because dimensional reduction of the 5D
$R^2$ gives the 4D $R^2$ term plus more terms involving 4D field
strengths. However we also see that the mismatch is subleading in
the expansion in $g_{top}$, and we can therefore conclude that the
5D $R^2$ term captures the subleading correction to the area law.

\section{Summary}
We have shown that higher dimension
corrections to the 5D effective action do give corrections to the black
hole/black ring entropy just as in 4D, but that the 5D situation is
currently under much less control than the 4D one.  Some leading
order computations were performed and found to give a partial match
between macroscopic and microscopic results. We hope these
computations will provide useful data for finishing the 5D
macro/micro story.

 \medskip
\section*{Acknowledgments}
\noindent We thank Michelle Cyrier, Davide Gaiotto  and Xi Yin
 for valuable discussions. This work was supported
in part by DOE grant DE-FG02-91ER40654.



\begin{thebibliography}{99}

\bibitem{bcov}
M.~Bershadsky, S.~Cecotti, H.~Ooguri, C.~Vafa, ``Kodaira-Spencer
Theory of Gravity and Exact Results for Quantum String Amplitudes'',
Commun.Math.Phys. \textbf{165} (1994) 311-428
[arXiv:hep-th/9309140];


I.~Antoniadis, E.~Gava, K.S.~Narain, T.R.~Taylor, ``Topological
Amplitudes in String Theory'', Nucl.Phys. B \textbf{413}
(1994)162-184 [arXiv:hep-th/9307158];


\bibitem{dutch}
G.L.~Cardoso, B.~de Wit, T.~Mohaupt, ``Corrections to macroscopic
supersymmetric black-hole entropy'', Phys.Lett. B \textbf{451}
(1999) 309-316 [arXiv:hep-th/9812082];

\bibitem{osv}
H.~Ooguri, A.~Strominger, C.~Vafa, ``Black Hole Attractors and the
Topological String'', Phys.Rev. D \textbf{70} (2004) 106007
[arXiv:hep-th/0405146];

\bibitem{gsybh}
D.~Gaiotto, A.~Strominger and X.~Yin, ``New Connections Between 4D
and 5D Black Holes''
 [arXiv:hep-th/0503217];


\bibitem{mohaupt}
T.~Mohaupt, ``Black Hole Entropy, Special Geometry and Strings'',
Fortsch.Phys. \textbf{49} (2001) 3-161 [arXiv:hep-th/0007195];


\bibitem{dlvp}
  B.~de Wit, P.~G.~Lauwers and A.~Van Proeyen,
 ``Lagrangians Of N=2 Supergravity - Matter Systems'', Nucl.\ Phys.\ B {\bf 255} (1985) 569;



\bibitem{berg}
  E.~Bergshoeff, S.~Cucu, T.~de Wit, J.~Gheerardyn, S.~Vandoren and A.~Van Proeyen,
  ``N = 2 supergravity in five dimensions revisited'',
  Class.\ Quant.\ Grav.\  {\bf 21}(2004) 3015-3042
  [arXiv:hep-th/0403045];


\bibitem{gutperl}
M.B.~Green, M.~Gutperle, ``Effects of D-instantons'', Nucl.Phys. B
\textbf{498} (1997) 195-227 [arXiv:hep-th/9701093];

\bibitem{chirita}
E.~Kiritsis, B.~Pioline, ``On $R^4$ threshold corrections in IIB
string theory and (p,q) string instantons'', Nucl.Phys. B
\textbf{508} (1997) 509-534 [arXiv:hep-th/9707018];

N.~Berkovits, C.~Vafa, ``Type IIB $R^4 H^{4g-4}$ Conjectures'',
Nucl.Phys. \textbf{B 533} (1998) 181-198 [arXiv:hep-th/9803145];


\bibitem{eemr}
H.~Elvang, R.~Emparan, D.~Mateos and H.~Reall, ``Supersymmetic black
rings and three-charge supertubes'', Phys.\ Rev.\ D {\bf 71}, 024033
(2005) [arXiv:hep-th/0408120];


\bibitem{bmpv}
J.~C.~Breckenridge, R.~C.~Myers, A.~W.~Peet and C.~Vafa, ``D-branes
and spinning black holes'', Phys.\ Lett.\ B {\bf 391}(1997) 93-98
[arXiv:hep-th/9602065];


\bibitem{cgms}
M.~Cyrier, M.~Guica, D.~Mateos and A.~Strominger, ``Microscopic
Entropy of the Black Ring'' [arXiv:hep-th/0411187];


\bibitem{ferrara}
I.~Antoniadis, S.~Ferrara, R.~Minasian, K.S.~Narain, ``$R^4$
Couplings in M and Type II Theories on Calabi-Yau Spaces'',
Nucl.Phys. B \textbf{507} (1997) 571-588   [arXiv:hep-th/9707013];

S.~Ferrara, R.~Khuri, R.~Minasian, ``M-Theory on a Calabi-Yau
Manifold'', Phys.Lett. B \textbf{375} (1996) 81-88
[arXiv:hep-th/9602102];

\bibitem{waldref}
R.M.~Wald, ``Black Hole Entropy is Noether Charge'',  Phys.Rev. D
\textbf{48} (1993) 3427-3431 [arXiv:gr-qc/9307038];

V.~Iyer, R.M.~Wald, ``Some Properties of Noether Charge and a
Proposal for Dynamical Black Hole Entropy'', Phys.Rev. D \textbf{50}
(1994) 846-864 [arXiv:gr-qc/9403028];

T.~Jacobson, G.~Kang, R.C.~Myers, ``Black Hole Entropy in Higher
Curvature Gravity'' [arXiv:gr-qc/9502009];


\bibitem{msw}
J.~Maldecena, A.~Strominger and E.~Witten, ``Black Hole Entropy in
M-Theory'', JHEP {\bf 9712} (1997) 002 [arXiv:hep-th/9711053];

\bibitem{myers}
R.~Myers, ``Black Holes in Higher Curvature Gravity'', Essays in
honour of C.V. Vishveshwara [arXiv:gr-qc/9811042];

\bibitem{kallars}
 R.~Kallosh, A.~Rajaraman, W.K.~Wong, ``Supersymmetric Rotating Black Holes and Attractors'', Phys.Rev. D \textbf{55} (1997) 3246-3249
  [arXiv:hep-th/9611094];

P.~Kraus, F.~Larsen, ``Attractors and Black Rings''
[arXiv:hep-th/0503219];


\bibitem{sventropy}
A.~Strominger and C.~Vafa, ``Microscopic Origin of the
Bekenstein-Hawking Entropy'', Phys.\ Lett.\ B {\bf 379} (1996) 99
[arXiv:hep-th/9601029].





\end{thebibliography}
\end{document}